%% file: maxent23.tex
\documentclass[nomathfonts]{aipproc} 
\layoutstyle{6s}
\usepackage{xspace,amsmath,amssymb,bbm,graphicx}
\input{alphabet}   
\input{abrege}    
\input{abrmath}

\input{beginend}

\title{An alternative inference tool to total probability formula and its applications}

\author{Adel Mohammadpour}
 {
 address={School of intelligent Systems, IPM, Tehran, Iran,\linebreak 
 {Present address: Laboratoire de Math\'ematique,
 Equipe Probabilit\'es, Satistiques et Mod\'elisation,
 Universit\'e de Paris-Sud, Batiment 425, 91405 Orsay,France.}\linebreak 
 {Permanent address: Department of Statistics,
 Faculty of Mathematics \& Computer  Science, Amirkabir University
 of Technology, 424 Hafez Ave., 15914 Tehran, Iran}\linebreak },
 email={\tt adel@aut.ac.ir}
 }
\author{Ali Mohammad-Djafari}
 {
  address = {Laboratoire des Signaux et Syst\`emes,\linebreak 
  Unit\'e mixte de recherche 8506 (CNRS-Sup\'elec-UPS) \linebreak  
  Sup\'elec, Plateau de Moulon, 91192 Gif-sur-Yvette, France},
  email = {djafari@lss.supelec.fr}
 }
\date{}

\newtheorem{theorem}{Theorem}
\newtheorem{definition}{Definition}
\newtheorem{proposition}{Proposition}
\newtheorem{example}{Example}
\newtheorem{lemma}{Lemma}[theorem]
\newtheorem{corolary}{Corollary}[proposition]

\def\d#1{\;\mbox{d}#1}

\def\tFx{\widetilde{F}_{X}}

\def\Fx{F_{X}(x)}
\def\Ftx{\widetilde{F}_{X}(x)}

\def\Fxnu{F_{X|\nu}(x|\nu)}
\def\Ftxnu{\widetilde{F}_{X|\nu}(x|\nu)}

\def\Fxtheta{F_{X|\theta}(x|\theta)}
\def\Ftxtheta{\widetilde{F}_{X|\theta}(x|\theta)}

\def\Fxnutheta{F_{X|\nu,\theta}(x|\nu,\theta)}

\def\fxtheta{f_{X|\theta}(x|\theta)}
\def\ftxtheta{\widetilde{f}_{X|\theta}(x|\theta)}

\def\Ftxm{\widetilde{F}_{X}(x_-)}
\def\Ftxp{\widetilde{F}_{X}(x_+)}

\def\Fxinu{F_{X|\nu}(x_i|\nu)}
\def\Ftxi{\widetilde{F}_{X}(x_i)}

\def\fxitheta{f_{X|\theta}(x_i|\theta)}
\def\ftxitheta{\widetilde{f}_{X|\theta}(x_i|\theta)}

\def\thetah{\widehat{\theta}}
\def\thetat{\widetilde{\theta}}

\begin{abstract}
An alternative inference tool  for using prior information to
calculate marginal distribution function in the Bayesian
statistics is suggested. A few applications of this new tool are
given.
\end{abstract}

\begin{document}
\maketitle

\section{Introduction}
Total probability and Bayes formula are two basic tools for using
prior information in the Bayesian statistics. In this paper we
introduce an alternative tool for using prior information. 
This new toold enables  us to improve some traditional results in
statistical inference. However, as far as the authors know, there
is no work on this subject, except [1]. The results of this paper
can be extended to other branches of probability and statistics.

In Section 2 total probability formula based on median is defined
and its basic properties are proved. A few applications of this new
tool are given in Section 3. 
All computations and plots are done
using the  S-PLUS\footnote{S-PLUS $\copyright$ 1988, 1999
MathSoft, Inc.} software system.

\section{Total probability formula based on median}
Let $X$ be a continuous
random variable with distribution function $\Fxnu$,
which depends on parameter $\nu$ with  known and continuous
density function $\pi(.)$.
 The
marginal distribution function  $X$ can be calculated by total
probability formula, i.e.

\beq \label{eq1}
\Fx=\int_{-\infty}^\infty
\Fxnu\;\pi(\nu) \d{\nu}
\eeq

Therefore $F_X$ is a
weighted mean of $F_{X|\nu}$, i.e., $F_{X}$ is the expected value
of $F_{X|\nu}$ over $\pi$. Our idea for the following definition
is
similar to (\ref{eq1}).

\begin{definition}
\label{def1}
Let $X$ have a distribution function
depending on parameter $\nu$, where $\nu$ has a density function
$\pi(.)$. The marginal distribution function of $X$ based on
median, $\Ftx$, is defined as the median of $F_{X|\nu}(x|v)$
over $\pi$.
\end{definition}

We recall that median is robust with respect to outlier data, but
mean is not. To simplify calculations of  $\Ftx$, we use
definition of median in statistics. That is we calculate 
$\Ftx$ by solving the following equation

\beq  \label{eq2}
F_{\Fxnu}(\Ftx)=\frac{1}{2},\;\;\mbox{~or equivalently~}\;\;
P(\Fxnu\leq \Ftx)=\frac{1}{2}.
\eeq

The following theorem states an important property of $\Ftx$.

\begin{theorem}
\label{theorem1}
$\Ftx$ is a non-decreasing and continuous function of $x$.
\end{theorem}
{\bf Proof:} Let $x_1<x_2$. For $i=1,2$, take
$$k_i=\Ftxi \mbox{~~~and~~~} Y_i=\Fxinu.$$ 
Then using~\ref{eq2} we have

$$P(Y_1\le k_1)=P(Y_2\le k_2)=\frac{1}{2}.$$ 
We also have 
$$ Y_1 \le Y_2.$$
Therefore,
$$P(Y_1\le k_1)=P(Y_2\le k_2) \le P(Y_1 \le k_2),$$
i.e. $k_1 \le k_2$ or equivalently $\Ftx$ is non-decreasing.
\vspace*{.7cm}
\\
If $\Ftx$ is a non-decreasing function, then
$$\Ftxm=\lim_{t\uparrow x}\tFx(t) \;\;\mbox{and}\; \;\tilde
F_X(x_+)=\lim_{t\downarrow x}\tFx(t)$$ {\em exist} and are
{\em finite} (e.g. [2]).

\medskip
Further, $\Fxnu$ is continuous with respect to $x$, and
so
$$ P(F_{X|\nu}(x_-|\nu)\leq \tilde
F_{X|\nu}(x_-)) = P(\Fxnu\leq \Ftxm),$$
$$
 P(F_{X|\nu}(x_+|\nu)\leq \Ftxp)
= P(\Fxnu\leq \Ftxp).  $$
And by (\ref{eq2}) we have  

\beqn
P(\Fxnu\leq \Ftxm) &=&
P(\Fxnu\leq \Ftx) \nonumber
\\ 
&=& P(\Fxnu\leq \Ftxp).
\label{eq3}
\eeqn
If
$Y=\Fxnu$ has an increasing distribution function, then
$$\Ftxm=\Ftx=\Ftxp$$   and by (\ref{eq3})
$\Ftx$ is {\em continuous}.

On the other hand $\Ftx$ is the median of $Y$, $0\le Y
\le 1$, and so  $$0\le \Ftx\le 1.$$
 Also, by Theorem~\ref{theorem1}, $\tFx(+\infty)$ and 
$\tFx(-\infty)$
 {\em exist}
as $$\lim_{t\uparrow +\infty}\tFx(t)\; \;\mbox{and} \;\;
\lim_{t\downarrow -\infty}\tFx(t)$$ respectively. Therefore
$\Ftx$ is a {\em distribution function} if $$\tilde
F_X(+\infty)=1 \;\;\mbox{and} \;\; \tFx(-\infty)=0.$$

\begin{example}
\label{ex1}
Let ${ X}$ be exponentially distributed,  i.e.
$$\Fxnu=1-e^{-\nu x},\;\; x>0$$
and assume  $\pi(\nu)=1,\;0<\nu\leq 1$. 
\end{example}
In this example we can
calculate $\tFx$ exactly by equation~(\ref{eq2}) as follows
\begin{eqnarray*}
&                     & P(1-e^{-\nu x} \leq \Ftx)=\frac{1}{2}\\
& \Longleftrightarrow & P(\nu\leq \frac{-1}{x}\ \ln (1-\Ftx))=\frac{1}{2}\\
& \Longleftrightarrow & \frac{-1}{x}\ \ln (1-\Ftx)=\frac{1}{2}\\
& \Longleftrightarrow & \Ftx=1-e^{-x/2},\;x>0.
\end{eqnarray*}
It can be shown that $\tFx$ is a distribution function.
Moreover,
$$\Fx=1+\frac{1}{x}(e^{-x}-1),\;x>0.$$

\medskip
In some problems we cannot calculate $\tFx$ exactly. But,
we can approximate it in the two following cases. \vspace*{.5cm}

\newpage
\noindent{\bf Algorithm M1:} 
When $\Fxnu$ has an analytic form,
but we cannot calculate $\tFx$ analytically.
\begin{enumerate}
\item Fix $x$ (say $x_0$)
\item Generate $K$ sample  for $\nu$ by using $\pi(\nu)$ (say
$\nu_k,\;k=1\cdots,K)$
\item Calculate the sample median of \\
$F_{X}(x_0;\nu_1),\cdots ,F_{X}(x_0;\nu_K)$
\item Repeat from step 1 with another choice of $x$.
\end{enumerate}

\medskip
\noindent{\bf Algorithm M2:}
When $\Fxnu$ has not an analytic form.
\begin{enumerate}
\item Fix $x$ (say $x_0$)
\item Generate $K$ sample for $\nu$ by using $\pi(\nu)$ (say
$\nu_k,\;k=1\cdots,K)$
\item Generate $L$ sample for  $X|\nu_k$ for each $k(=1,\cdots,K)$ \\ (say $(x_1|\nu_1,\cdots,x_L|\nu_1)
\cdots (x_1|\nu_K,\cdots,x_L|\nu_K) $)
\item Calculate the empirical distribution function of $X|\nu_k$ for each $k(=1,\cdots,K)$ (based on
generated samples in the previous step)
\item Calculate the sample median of empirical
distribution functions in  step~4
and repeat from step 1 with another choice of $x$.
\end{enumerate}

\begin{figure}[hp]
\btabu{c}
\includegraphics[width=\textwidth]{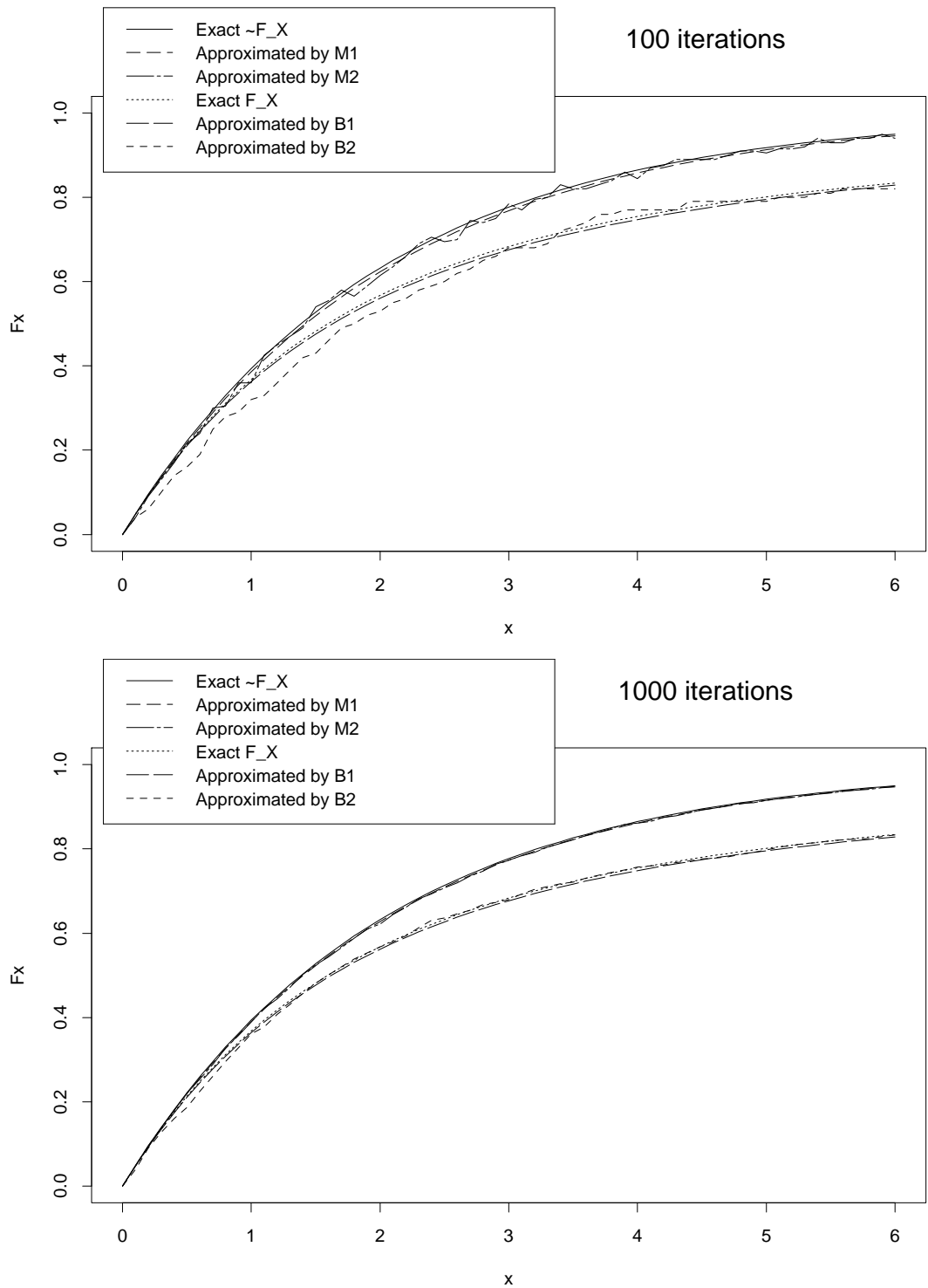}
\etabu
\caption{Graphs of $\tFx$, $F_X$, and their
approximations in Example~\ref{ex1} for $K=100,1000$}
\label{fig1}
\end{figure}

\medskip
\noindent{\bf Remark 1} 
We can approximate $F_X$ by algorithms similar
to M1 and M2  (are called  B1 and B2 corresponding to M1 and M2).
\vspace*{.5cm}
\\
Figure~\ref{fig1}, shows the graphs of $\tFx$, $F_X$, and their
approximations for Example~\ref{ex1}.

\begin{figure}[hp]
\btabu{c}
\includegraphics[width=\textwidth]{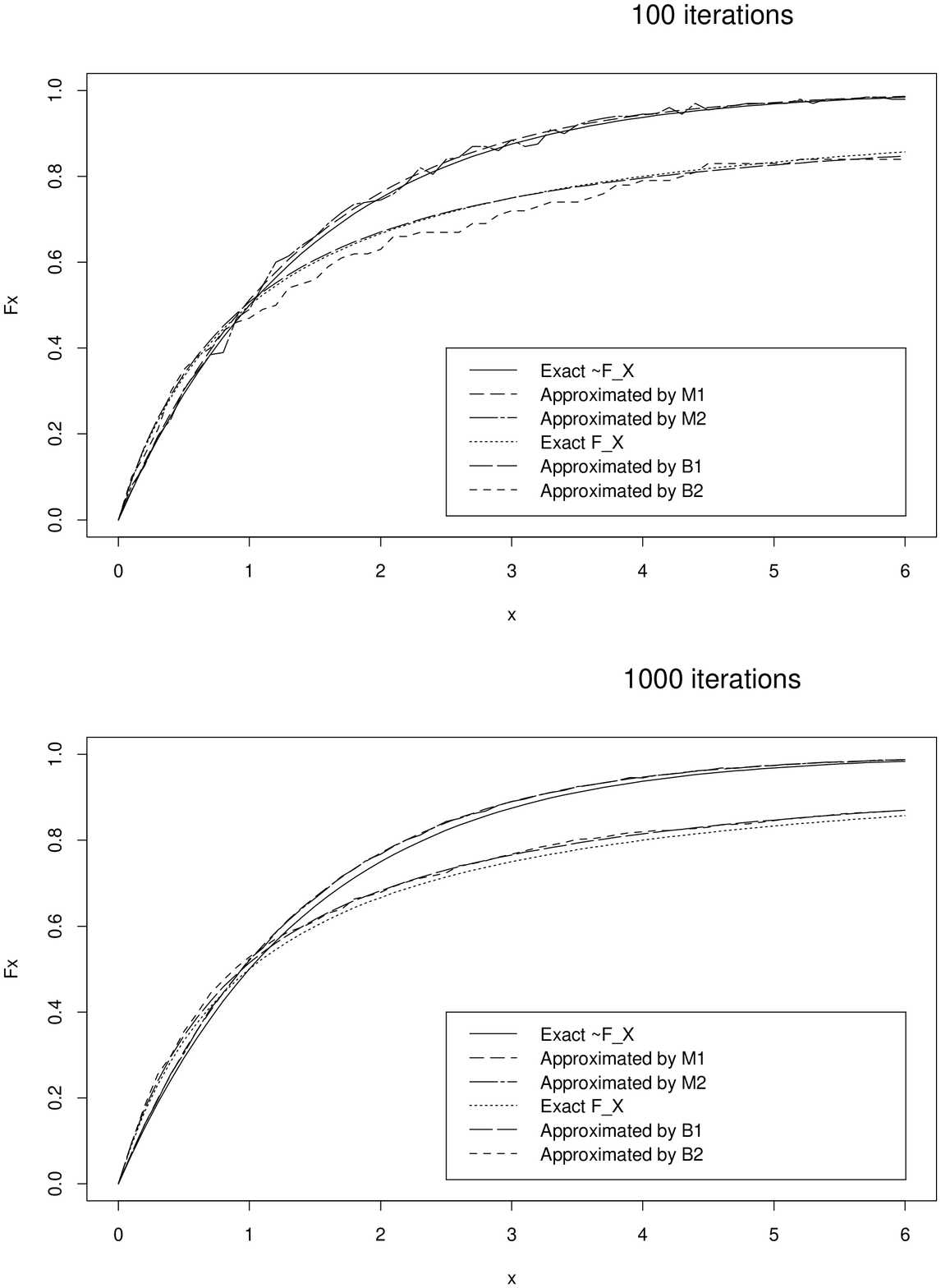}
\etabu
\caption {Graphs of $\tFx$, $F_X$, and their
approximations in Example~\ref{ex2} for $K=100,1000$}
\label{fig2}
\end{figure}

\medskip
\begin{example}
\label{ex2}
Let ${ X}$  be exponentially distributed,
(similar Example~\ref{ex1}) i.e.
$$\Fxnu=1-e^{-\nu x},\;\; x>0$$
but here  $$\pi(\nu)=e^{-\nu}, \; \nu>0.$$ 
\end{example}
In this case
$$\Ftx=1-e^{x\ln(1/2)},\;x>0,$$
is a distribution function and
$$\Fx=1-\frac{1}{x+1},\;x>0.$$
 Figure~\ref{fig2} shows their graphs.

\newpage
\section{Hypothesis Testing}

In this section we introduce a few applications of $\tFx$
to improve traditional results in statistical inference.

In the previous section we showed that $\tFx$ is a
distribution function under a  few conditions.
 If $\tFx$ depends on some
unknown parameters we can apply classical methods in statistics
to make inference about the unknown parameters. For example,
uniformly most powerful (UMP) test can be calculated by
Karlin-Robin theorem \cite{}, or the most powerful (MP) test can be
calculated by the following version of Neyman-Pearson lemma's.

\begin{lemma} \label{lem1}
Consider testing

\beq \label{eq4}
\left \{
   {{  H_0 : \theta =\theta_0} \atop
     { H_1 : \theta=\theta_1 }}
\right. ,
\eeq
where $\theta$ is an unknown parameter of $\tFx$
 and $\theta_0$, $\theta_1$  are fixed known numbers. 
If $\tFx$ does not depend on any other  unknown parameters under $H_0$
and $H_1$, then

\beq \label{eq5}
\phi(x)=
 \left \{
  {{  1 \;\;\;  \frac{d\Ftx}{dx}\mid_{\theta =\theta_1}
  > k \frac{d\Ftx}{dx}\mid_{\theta =\theta_0}}
  \atop
    { 0  \;\;\;
 \frac{d\Ftx}{dx}\mid_{\theta =\theta_1}
< k \frac{d\Ftx}{dx}\mid_{\theta =\theta_0}
    }}
\right. ,
\eeq
for some $k \geq 0$, is the MP test of its size for testing.
\end{lemma}
{\bf Proof:} Let   $ \Ftx= \frac{d\Ftx}{dx} $.
Then $ \Ftx$ is a continuous density function which does
not depend on any other unknown parameters under $H_0$ and
$H_1$. Therefore by the Neyman-Pearson lemma (\ref{lem1}) is the MP test
of its size for testing~(\ref{eq5}). 

\begin{example}
\label{ex3}
Consider testing 
\beq \label{eq6}
\left \{
   {{  H_0 : \mu =0} \atop
     { H_1 : \mu <0}}
\right. . 
\eeq
 
based on an observation from a normal
distribution $X\sim N(\mu,\sigma^2)$. 
\end{example}
If $\sigma^2$ is known, then
the family of normal distribution has Monotone Likelihood
Ratio (MLR) property and according to the Karlin-Robin theorem
\beq \label{eq7}
\phi_{\sigma^2} (x) = \left \{ {{ 1 \;\;\;\;\;\;\;
\frac{x}{\sigma}< z_{\alpha} }\atop { 0 \;\;\;\;\;\;\;
\frac{x}{\sigma}> z_{\alpha}} }\right. ,
\eeq 
is the Uniformly Most Powerful (UMP), of size $\alpha$ test function for
\ref{eq6}, where $P(Z<z_{\alpha})=\alpha $ and $Z\sim N(0,1)$. But, if
$\sigma^2$ is unknown, then the best test does not exist.

\begin{figure}[hp]
\btabu{c}
\includegraphics[width=\textwidth]{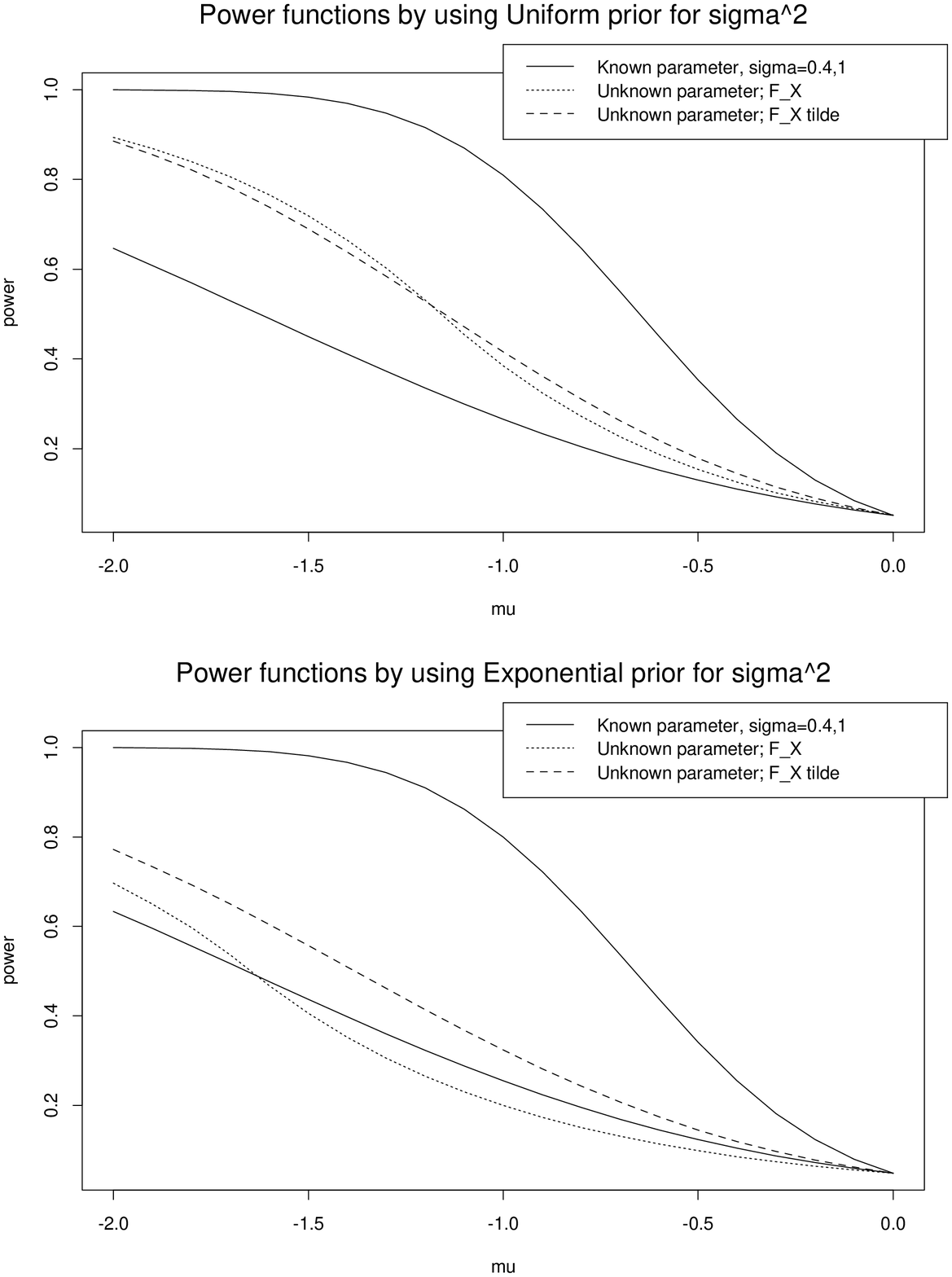}
\etabu 
\caption{The graphs of power functions when the variance
has a uniform and exponential prior. }
\label{fig3}
\end{figure}

In the Bayesian approach, where   $\sigma^2$ (variance)  has a
prior  density function such as uniform or exponential (where
defined in Examples~\ref{ex1} and ~\ref{ex2}
 respectively), we can find  marginal distribution functions
  $F_X$ and $\tFx$ which  depends on $\mu$.
Figure~\ref{fig3} shows the graphs of  power functions of the
tests based on $F_X$ and $\tFx$ for $\alpha=0.05$. We also
plot the graph of power function of \ref{eq7} for $\sigma=0.4,1$
(i.e. when $\sigma$ is known). The graphs show that the test
based on $\tFx$  is better than the test
based on $F_X$, when we use exponential prior for $\sigma^2$.

Moreover, we plot the graphs of power functions of the tests
based on $F_X$ and $\tFx$ in the two cases of 
 uniform and  exponential  prior distributions for
$\sigma$ (standard deviation)
 in Figure~\ref{fig3}. The result is incredible! The test
based on $\tFx$ is much better than the test based on $F_X$.

\begin{figure}[hp]
\btabu{c}
\includegraphics[width=\textwidth]{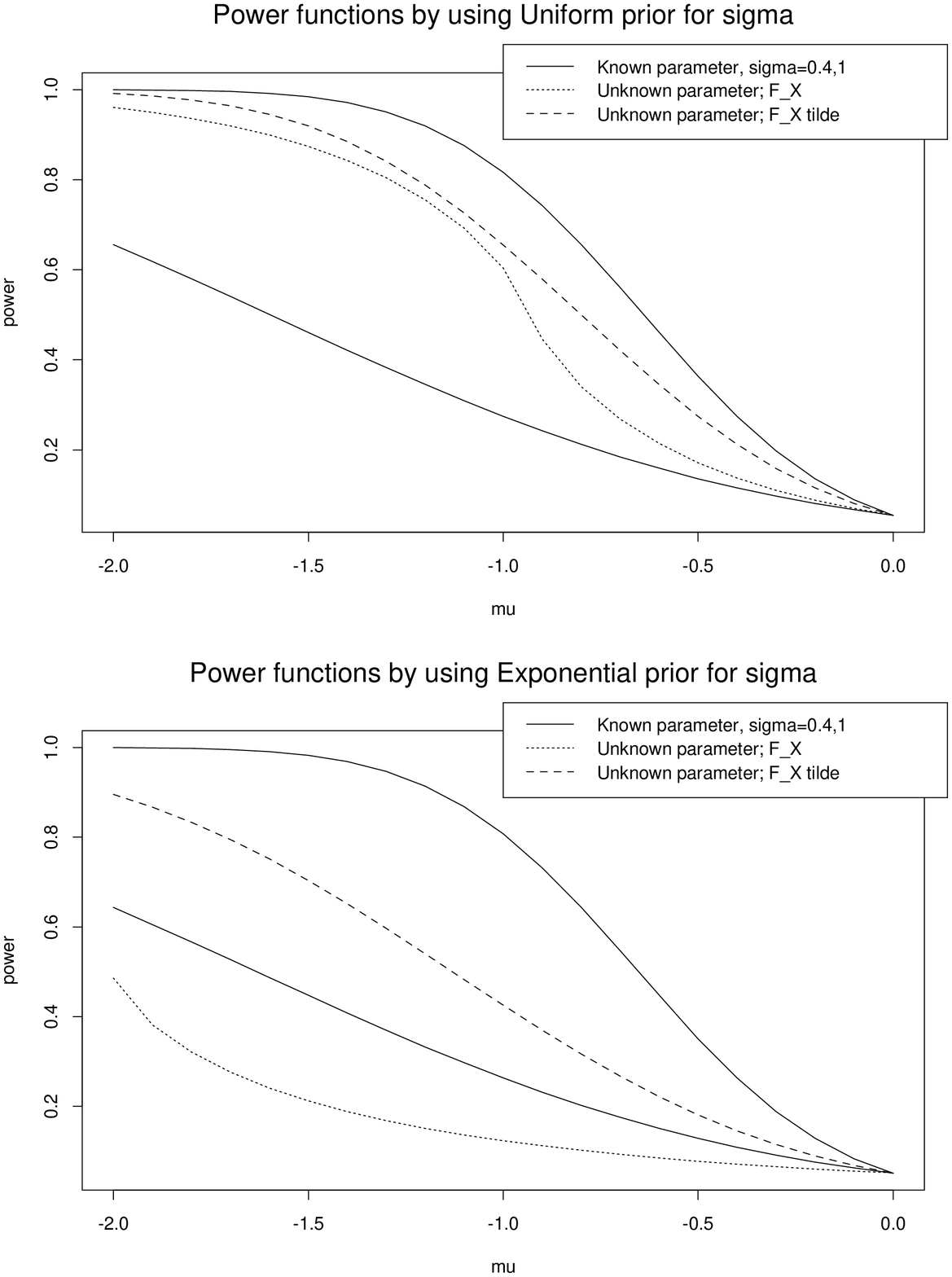}
\etabu 
\caption{The graphs of power functions, when the
standard deviation has a uniform and exponential prior. }
\label{fig4}
\end{figure}

\section{Parameter estimation}

Consider the case where the distribution of X depends on two parameters 
$\theta$ and $\nu$, \ie, we have $\Fxnutheta$. Then, we can define 
$\Fxtheta$ and $\Ftxtheta$ as in previous case. Then, by derivating them with respect to $x$, we can also define $\fxtheta$ and $\ftxtheta$. 
Assume now that we have a data set $x-1, \cdots, x_N$ where we assume its 
distribution to be $\Fxnutheta$ and where we have prior knowledge $\pi(\nu)$, 
and we want to estimate $\theta$ for this data set. 

The classical MLE is defined by
\beq
\thetah=\argmax_{\theta}\left\{L(\theta)=\prod_{i=1}^N \fxitheta\right\}
\eeq
Similarly, based on our new criterion we propose the following
\beq
\thetah=\argmax_{\theta}\left\{\widetilde{L}(\theta)=\prod_{i=1}^N \ftxitheta\right\}.
\eeq
To show the relative performances of these two estimators, we

TO COMPLETE LATER 

\section{Conclusion}
We introduced an alternative inference tool for using prior information $\pi(\nu)$ by defining a marginal function $\Ftxnu$ which is based on median in place of $\Fxnu$ which is the expected value of $\Ftxnu$ with respect to $\pi(\nu)$. 
We proved that $\Ftxnu$ is a non-decreasing and continuous function of $x$ and presented some of its applications and its performances in hypothesis testing and in parameter estimation.

\end{document}

%% file: abrmath.tex
\newsavebox{\fminibox}
\newlength{\fminilength}
\newenvironment{fminipage}[1][\linewidth]
  {\setlength{\fminilength}{#1}
   \begin{lrbox}{\fminibox}\begin{minipage}{\fminilength}}
  {\end{minipage}\end{lrbox}\noindent\fbox{\usebox{\fminibox}}}

  \def\+{^\dagger}

\def\nequiv{\not\kern-.05em\equiv}
\def\egal{\kern-.5em=\kern-.5em}        
\def\propt{\kern-.2em\propto\kern-.2em} 

\def\argmax{\mathop{\mathrm{arg\,max}}} 

\def\intdouble{\int\kern-0.3em\int}
\def\inttriple{\int\kern-0.3em\int\kern-0.3em\int}

\def\rond#1{\overset{\kern-0.33em~_\circ}{#1}}
\def\rondit[#1]#2{\overset{\kern#1~_\circ}{#2}}


%% file: beginend.tex
\def\babs{\begin{abstract}}             \def\eabs{\end{abstract}}
\def\barr{\begin{array}}                \def\earr{\end{array}}
\def\bcc{\begin{center}}                \def\ecc{\end{center}}

\def\bdes{\begin{description}}          \def\edes{\end{description}}
\def\bdoc{\begin{document}}             \def\edoc{\end{document}}
\def\ben{\begin{enumerate}}             \def\een{\end{enumerate}}
\def\beqn{\begin{eqnarray}}             \def\eeqn{\end{eqnarray}}
\def\beqnl#1{\beqn\label{#1}}           \def\eeqnl#1{\label{#1}\eeqn}
\def\beqnx{\begin{eqnarray*}}           \def\eeqnx{\end{eqnarray*}}
\def\bseqn{\begin{subeqnarray}}         \def\eseqn{\end{subeqnarray}}
\def\beq#1\eeq{\begin{equation}#1\end{equation}}
\def\bal#1\eal{\begin{align}#1\end{align}}
\def\balx#1\ealx{\begin{align*}#1\end{align*}}
\def\beqx{$$}                           \def\eeqx{$$}
\def\bfig{\protect\begin{figure}}       \def\efig{\protect\end{figure}}
\def\bfigx{\protect\begin{figure*}}     \def\efigx{\protect\end{figure*}}
\def\bfigt{\protect\begin{figurette}}   \def\efigt{\protect\end{figurette}}
\def\bfl{\begin{flushleft}}             \def\efl{\end{flushleft}}
\def\bfr{\begin{flushright}}            \def\efr{\end{flushright}}
\def\bit{\begin{itemize}}               \def\eit{\end{itemize}}
\def\bmi{\begin{minipage}}              \def\emi{\end{minipage}}
\def\bfmi{\begin{fminipage}}            \def\efmi{\end{fminipage}}
\def\bpic{\begin{picture}}              \def\epic{\end{picture}}
\def\bqu{\begin{quote}}                 \def\equ{\end{quote}}
\def\bqun{\begin{quotation}}            \def\equn{\end{quotation}}
\def\bsl{\begin{slide}}                 \def\esl{\end{slide}}
\def\btabb{\begin{tabbing}}             \def\etabb{\end{tabbing}}
\def\btabl{\begin{table}}               \def\etabl{\end{table}}
\def\btablx{\begin{table*}}             \def\etablx{\end{table*}}
\def\btab{\begin{tabular}} 
\def\btabu{\begin{tabular}}             \def\etabu{\end{tabular}}
\def\btabx{\begin{tabular*}}            \def\etabx{\end{tabular*}}
\def\bbib{\begin{thebibliography}{999}} \def\ebib{\end{thebibliography}}
\def\bver{\begin{verbatim}}             \def\ever{\end{verbatim}}
\def\bca{\begin{cases}}                          \def\eca{\end{cases}}